\documentclass[aps,pra,superscriptaddress,twocolumn]{revtex4-1}
\usepackage[utf8]{inputenc}
\usepackage[english]{babel}
\usepackage{amsmath}
\usepackage{amsfonts}
\usepackage{amssymb}
\usepackage{graphicx}
\usepackage{amsthm}
\usepackage{color}
\usepackage{hyperref}
\usepackage{paralist}
\newcommand{\ket}[1]{\ensuremath{|#1\rangle}}
\newcommand{\bra}[1]{\ensuremath{\langle #1|}}

\newcommand{\dbraket}[2]{\ensuremath{\langle #1 | #2 \rangle}}
\newcommand{\ketbra}[1]{\ensuremath{| #1 \rangle\langle #1 |}}
\newcommand{\dketbra}[2]{\ensuremath{| #1 \rangle\langle #2 |}}
\newcommand{\norm}[1]{\ensuremath{\left|#1\right|}}

\begin{document}
\title{Coherent-state discrimination  via non-heralded probabilistic amplification} 
\author{Matteo Rosati}
\affiliation{NEST, Scuola Normale Superiore and Istituto Nanoscienze-CNR, I-56127 Pisa,
Italy.}
\author{Andrea Mari} 
\affiliation{NEST, Scuola Normale Superiore and Istituto Nanoscienze-CNR, I-56127 Pisa,
Italy.}
\author{Vittorio Giovannetti} 
\affiliation{NEST, Scuola Normale Superiore and Istituto Nanoscienze-CNR, I-56127 Pisa,
Italy.}

\begin{abstract}
A scheme for the detection of low-intensity optical coherent signals was studied which uses a 
probabilistic amplifier operated in the non-heralded version, as 
the underlying non-linear operation to improve the detection efficiency. This approach allows us to improve the statistics by keeping track of  all possible outcomes of the amplification stage (including failures). When compared with an optimized Kennedy receiver, the resulting discrimination success probability we obtain presents a gain up to $\sim 1.85\%$ and it approaches the Helstrom bound appreciably faster than the Dolinar receiver, when employed in an adaptive strategy. We also notice that the advantages obtained can be ultimately 
associated with the fact that, in the high gain limit, the non-heralded version of the probabilistic amplifier induces a partial dephasing which preserves quantum coherence among low energy eigenvectors while removing it elsewhere. A proposal to realize such transformation based 
on an optical cavity implementation is presented.
\end{abstract}
\maketitle

\section{Introduction}
In quantum mechanics it is impossible to perfectly distinguish two or more non-orthogonal states. This basic observation has led to the study of minimum-error discrimination \cite{QSDiscriminationRev}, i.e. the technique of discriminating a set of quantum states with the lowest error allowed by the laws of physics. Following the pioneering work of Helstrom \cite{Hel}, who provided in particular a lower bound to the error probability in the two-state discrimination problem, research has focused on achieving that bound in practice \cite{Ken,OpKen,Dol,DolMultiplexed,Marq,AdaptiveLOCC,Multicopy,OpMeas2,QKD,ImplProj,ImplProj2,OpGauss,OpPrior,SeqNull,DolExp,DolExp2}. This task has key importance in optical communication, which is usually modeled by considering weak coherent states that encode binary information in phase or amplitude modulation \cite{RevGauss}. Such states are largely overlapping at low intensity, as it happens with fiber or free-space communication, and it is thus extremely important to design receivers that discriminate them as efficiently as possible, in order to reach the ultimate quantum bound on communication capacity \cite{HolBound,LossyBos}. 
It turns out that the optimal theoretical measurement achieving the Helstrom bound in this case 
 is  highly non-linear~\cite{OpMeas}  and practically impossible to implement with current technology.
A first realistic, yet sub-optimal, receiver  was proposed by Kennedy \cite{Ken}: it employs a coherent displacement operation that perfectly nulls one of the two possible signals (or the most favored one if they are not equiprobable), followed by photon detection. If the detector registers no photon, the result is interpreted as successful identification of the nulled signal, else if one or more photons are detected, the other signal is chosen. This receiver captures the main ingredient in coherent-state discrimination, i.e. signal nulling, which restricts the source of errors only to the overlap of the non-nulled signal(s) with the vacuum. Nevertheless at low intensity values its performance is lower than that of conventional receivers based on homodyne or heterodyne measurements; in particular Takeoka and Sasaki \cite{OpGauss} proved that the homodyne receiver is optimal among all possible Gaussian measurements.
Better results, which surpass the homodyne detection also at low intensity values, can be obtained by employing an imperfect nulling technique (or optimized Kennedy scheme), where the displacement  of the Kennedy scheme is chosen so as to maximize the success probability of the protocol, i.e. the difference between the vacuum-overlaps of the two states~\cite{OpKen,Marq}.  Further improvements can finally be obtained by embedding the above techniques into a multiplexing procedure~ \cite{ImplProj,ImplProj2,DolMultiplexed,AdaptiveLOCC,Multicopy,OpMeas2}  along the line first suggested by Dolinar~\cite{Dol}: here  the received coherent signal gets first split in $N$  lower-intensity copies that are then  individually probed (say via an optimized Kennedy detection) in a feedforward-adaptive routine where the settings of the forthcoming detectors are determined by the outcomes of previous ones. This approach ensures a reduction of the error probability as $N$ increases to the extent to which, assuming perfect photo-counters, it allows for the saturation of the Helstrom bound in the asymptotic limit of infinitely many iterations.

 In the present paper we analyze a  detection method (see Fig.~\ref{contourplot}a) which potentially could outperform the optimized Kennedy scheme by relying on a special class of non-linear effects 
 that originate from the action of the Probabilistic Amplifier (P-Amp) proposed recently by Ralph and Lund \cite{RalphStart}; more specifically in our analysis we employ the optimal theoretical description of such device presented in ~\cite{Caves,OptimalNLA}.
 We remind  that  a P-Amp of gain $g\geq1$  performs  probabilistically  the amplification of a coherent state $\ket{\alpha}$ into $\sim\ket{g\alpha}$, the failure events being highly probable but heralded by a triggering signal which allows one to discard them ~\cite{RalphStart,AltAmp,ADagA,PhaseAmp,ExpRalph,ExpSciss,ExpZav,ExpSciss2,ExpQKD,AmpAsSq,CompareSqAmp,AmpSymp,Caves,OptimalNLA}. 
To improve the statistics we operate such device by considering a non-heralded version (nh-P-Amp in brief) of the scheme  presented in Refs.~\cite{Caves,OptimalNLA}, i.e. 
we  act on the incoming signal with a standard P-Amp machine 
with the only difference that all events, failures included,  are accepted in the subsequent stages of the detection process. 
Specifically
starting from the input coherent states $\ket{\pm\alpha}$ we have to discriminate, we first apply a perfectly-nulling displacement for the favored one between them. Next we make use of the nh-P-Amp (see Sec.~\ref{sec2}).
 In this way the device acts as a completely positive and trace-preserving (CPTP) map which leaves the nulled state in the vacuum, while sending the second state into a mixture of the target amplified state, which is farther away from the vacuum, and of a complex truncated state, being neither the original coherent state nor its desired amplified version. By applying a final displacement operation which we optimize in order to maximize the success probability, we show 
that the resulting detection scheme surpasses the optimized Kennedy one 
 for any value of the amplifier's gain, reaching its optimal working regime for sufficiently high values of the gain.
A limiting factor of our proposal is the fact that in all implementations of the P-Amp discussed so far ~\cite{ExpRalph,ExpSciss,ExpZav,ExpSciss2,ExpQKD}, 
the attained values of  $g$ are relatively small. This problem can however be overcome by noticing that in the
 $g\rightarrow\infty$ limit, the nh-P-Amp we study actually becomes a partial dephasing operation, which destroys any coherence between the zero and one-photon subspace and the rest.
While typically one would be tempted to consider such dephasing as noise and hence detrimental, quite surprisingly it turns out to be 
a key ingredient for the success of the proposed scheme, contributing to  effectively reduce the overlap between the unfavored state and the optimized displaced vacuum.
 Motivated by this observation 
   in Sec.~\ref{sec3} we study a modified detection scheme where the  nh-P-Amp is substituted by a simpler dephaser, which preserves coherence only in the zero and one-photon subspace. The results being still positive in comparison with the optimized Kennedy detection, we eventually provide a possible implementation of this dephaser, which makes use of a cavity-atom system at resonance.
Finally we analyze the performance of the partial dephaser before a general active gaussian transformation (Sec.~\ref{sec4}) and in the Dolinar scheme  (Sec.~\ref{sec5}), reporting  improvements also in these cases. The paper ends with Sec.~\ref{sec6}, where we draw some conclusions, and with a couple of Appendices devoted to illustrate some technical points raised in the main text.

\section{The nh-P-Amp receiver}\label{sec2}
 The non-heralded version of  a P-Amp of gain $g\geq 1$ and cutoff $n$  we analyze here can be described as   a CPTP map $\mathcal{A}_{g,n}$ characterized by two Kraus operators\begin{align}
&\hat{M}_{S}=g^{-n}\sum_{k=0}^{n}g^{k}\ketbra{k}+\sum_{k=n+1}^{\infty}\ketbra{k},\\ 
 &\hat{M}_{F}=\sum_{k=0}^{n}\left(1-g^{-2(n-k)}\right)^{1/2}\ketbra{k},
 \end{align}
which, according to the analysis presented in Refs.~\cite{Caves,OptimalNLA}, identify respectively success and failure in the amplification of a regular P-Amp (the vectors $|k\rangle$ being elements of the Fock basis). We stress that this optimal theoretical form of the P-Amp differs from those based  on conditional gaussian operations~\cite{ExpRalph,ExpSciss,ExpZav,ExpSciss2}  which, when employed in a non-heralded way, would result in a gaussian measurement and thus would perform certainly worse than homodyne detection~\cite{OpGauss}.

 \begin{figure}[t]
\includegraphics[scale=.3]{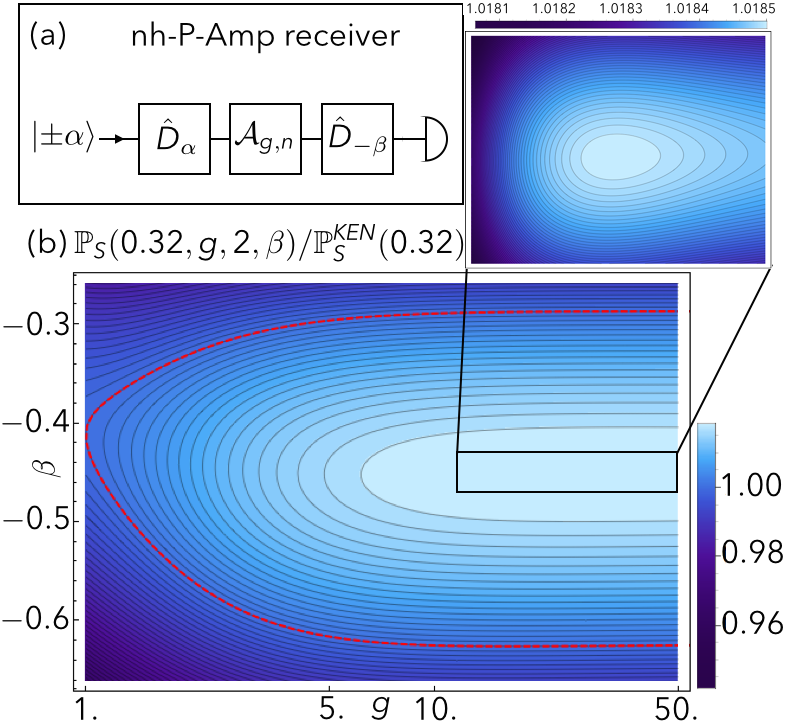}
\caption{(a) Schematic depiction of the nh-P-Amp receiver, where $\hat{D}_{\alpha}$ is the first coherent-state displacement, $\mathcal{A}_{g,n}$ is the non-linear nh-P-Amp and $\hat{D}_{-\beta}$ is the final optimized displacement, followed by an ordinary photon detector. The optimized Kennedy scheme is obtained by setting $g=1$, which amounts to removing $\mathcal{A}_{g,n}$. Other schemes described in the text rely on the substitution of $\mathcal{A}_{g,n}$ with simplified versions and/or on the introduction of an optimized squeezing operation $\hat{S}_{r}$ before the final displacement. (b) (Colour online) Contour plot of the ratio ${\mathbb P}_{{\bf S}}(\alpha,g,n,\beta)/{\mathbb P}_{{\bf S},opt}^{KEN}(\alpha)$ between the 
success probabilities~(\ref{succprobAmp}) of the nh-P-Amp receiver and the one associated  with the optimal Kennedy scheme,  for $n=2$, as a function of the gain $g$ and of the displacement $\beta<0$ for input intensity
 $\alpha=0.32$ (the optimal value of displacement for the Kennedy scheme being  $\beta^{op}_{Ken}\simeq 0.412$).  The red dashed line indicates  the points for which ${\mathbb P}_{{\bf S}}(\alpha,g,n,\beta)={\mathbb P}_{{\bf S},opt}^{KEN}(\alpha)$. The inset shows the optimal region for the nh-P-Amp receiver, in the range $\beta\in[-0.47,-0.43]$, $g\in[15,100]$.
 }\label{contourplot}
\end{figure}
\begin{figure}
\includegraphics[scale=.27]{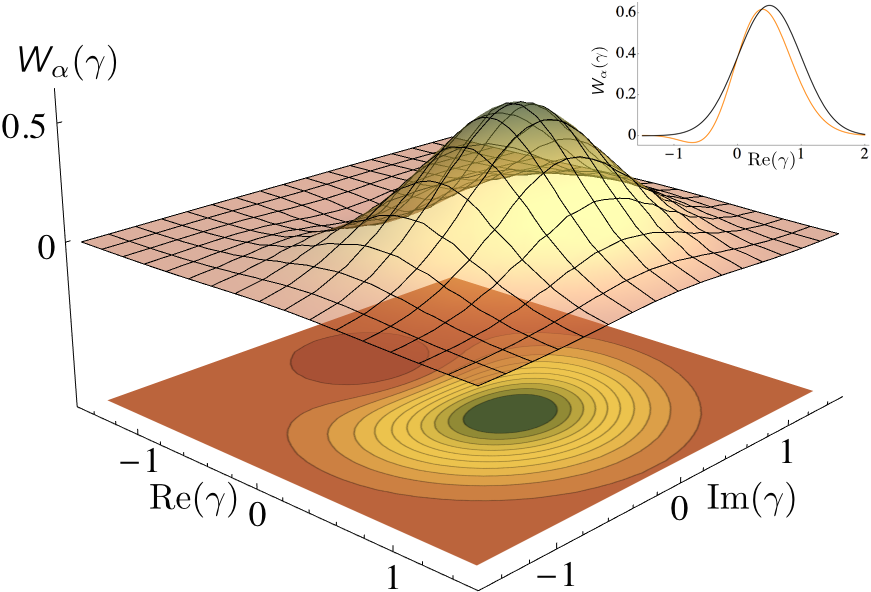}
\caption{(Colour online) Plot of the Wigner function $W_{\alpha}(\gamma)$ of the transformed coherent state $\mathcal{A}_{g,n}(\ketbra{\alpha})$ under the action of the nh-P-Amp. The inset shows a cut at $\operatorname{Im}(\gamma)=0$ of the same function (orange/light gray curve) and of the Wigner function of the input state $\ketbra{\alpha}$ (black curve). The former is far from an amplified version of the latter, still it exhibits a non-gaussian profile, which reduces   the overlap~(\ref{OVERLAP}) improving the scheme success probability~(\ref{succprobAmp}). Both plots are drawn at the optimal gain value of the nh-P-Amp receiver for the chosen intensity: $\alpha\simeq 0.5$, $g\simeq 28$, $n=2$.
 }\label{Wigner}
\end{figure}

Suppose hence we want to discriminate between the coherent states $\ket{\alpha_{0}}=\ket{-\alpha}$ and $\ket{\alpha_{1}}=\ket{+\alpha}$ produced respectively with prior probabilities $q_0$ and $q_1=1-q_0$, the former being favored (i.e. $q_0 \geq q_1$). As anticipated, in  our decoding scheme (Fig.~\ref{contourplot}a) we first apply a displacement which nulls  the favored state, then a nh-P-Amp transformation 
$\mathcal{A}_{g,n}$, 
and  finally a further displacement of $-\beta$, to be optimized later on (the latter being fundamental to get an improvement with respect to 
the Kennedy strategy, see Appendix~\ref{APPA} for details). Accordingly the states which enter the  photo-detector are 
\begin{eqnarray}
&&\ket{\alpha_{0}}\rightarrow\hat{D}_{-\beta}\mathcal{A}_{g,n}(\ketbra{0})\hat{D}_{-\beta}^{\dag}=\hat{D}_{-\beta}\ketbra{0}\hat{D}_{-\beta}^{\dag}, \\
&&\ket{\alpha_{1}}\rightarrow\hat{D}_{-\beta}\mathcal{A}_{g,n}(\ketbra{2\alpha})\hat{D}_{-\beta}^{\dag}, 
\end{eqnarray}
where in  the first term we used the fact that $\mathcal{A}_{g,n}$ leaves invariant the vacuum.
We now associate the event where no photons are detected to the arrival of $|\alpha_0\rangle$, while the others to the arrival of $|\alpha_1\rangle$. The success probability of the  protocol reads hence
\begin{equation}
{\mathbb P}_{{\bf S}}(\alpha,g,n,\beta)=q_{0}P(0|\alpha_{0})+q_{1}(1-P(0|\alpha_{1})),\label{succprobAmp}
\end{equation}
where   $P(0|\alpha_{k})$ is the probability of detecting no photons from the transformed final state associated with the input $|\alpha_k\rangle$. They can be expressed as 
 \begin{eqnarray} 
P(0|\alpha_{0})&=& | \bra{\beta}{0}\rangle |^2\;,\label{OVERLAP1}\\
 P(0|\alpha_{1})&=&\langle \beta | \mathcal{A}_{g,n}(\ketbra{2\alpha})|\beta\rangle\;, \label{OVERLAP}
 \end{eqnarray} 
respectively with 
$|\beta\rangle$ being the coherent state of
amplitude~$\beta$.  The action of $\mathcal{A}_{g,n}$ on an input coherent state of non-vanishing amplitude
 is depicted in Fig.~\ref{Wigner}, in terms of the Wigner function of the output state. We notice that the state has a slightly non-gaussian form:
  ultimately this is the key feature which allows one to
 improve the success probability of the scheme by reducing the overlap of 
 $\mathcal{A}_{g,n}(\ketbra{2\alpha})$ with the displaced vacuum $|\beta\rangle$.

Equation~(\ref{succprobAmp})  has to be compared  with  the standard Kennedy result, whose 
success probability can be recovered  from the same expression by simply setting $g=1$ (no amplification), i.e. 
\begin{eqnarray}
{\mathbb P}_{{\bf S}}^{KEN}(\alpha,\beta) = {\mathbb P}_{{\bf S}}(\alpha,g=1,n,\beta).
\end{eqnarray} 
Now, for a fixed value  of the input amplitude $\alpha$, we may optimize 
 \eqref{succprobAmp} with respect to three parameters: the amplifier's gain $g$, its internal cutoff degree $n$ and the displacement $-\beta$. 
  For $n=1$ and negative $\beta$ the success probability is a decreasing function of the gain, thus the optimal choice is not to amplify and the scheme simply resorts to the optimized Kennedy detector. 
  On the contrary,  for $n=2$ there is a whole range of non-trivial values of the gain which increase the success probability~(\ref{succprobAmp}) of the nh-P-Amp detector above that of the optimized Kennedy detector, i.e. above the value 
  ${\mathbb P}_{{\bf S},opt}^{KEN}(\alpha) =\max_{\beta} {\mathbb P}_{{\bf S}}^{KEN}(\alpha,\beta)$.
 In particular, at $\alpha\simeq 0.32$, already for $g=3$ we have an increase of $\sim1.26\%$ with respect to ${\mathbb P}_{{\bf S},opt}^{KEN}(\alpha)$ (the maximum increase of $\sim 1.85\%$  is attained for $g\sim 31$, while for $g\rightarrow\infty$ it lowers to $\sim 1.84\%$),  see  
  Fig.~\ref{contourplot}b. 
 \section{Partial dephaser receiver}\label{sec3}
It is of primary importance to stress that 
  replacing  $\mathcal{A}_{g,n}$  with an ordinary parametric amplifier or, more generally, a phase-insensitive gaussian channel would not provide
  the advantages  reported in the previous section: in this case in fact the success probabilities would be worse than those attainable with the optimized Kennedy detector   (see Appendix~\ref{APPB}). Therefore the mere amplification of the incoming signals cannot account for the improvement of the performances. 
Still, by numerical analysis of the case $n=2$,  we observe that, fixing $\beta$ and $\alpha$,   the advantage gained from the application of $\mathcal{A}_{g,2}$ is 
 almost optimal in the infinite-gain limit, i.e. for $g\rightarrow\infty$ (see Fig.~\ref{contourplot}b).

  In this regime the Kraus operators which define the action of the nh-P-Amp reduce to simple projectors 
  on the subspace of $2$ or more photons ($\hat{M}_{S}\rightarrow \hat{P}_{\geq 2}$)  and on its complementary subspace ($\hat{M}_{F}\rightarrow \hat{P}_{< 2}$). Accordingly 
 $\mathcal{A}_{\infty,2}$ reduces to a partial dephasing channel that selectively removes the coherence among  such subspaces while preserving any other form of quantum coherence in the system. 
 This observation
   along with the fact that, to our knowledge, the nh-P-Amp in the form \cite{Caves,OptimalNLA} has not been experimentally demonstrated yet, 
   leads us to replace $\mathcal{A}_{\infty,2}$ of Fig.~\ref{contourplot} with a simplified version of such a map
    for which we propose a possible implementation. 
Specifically we consider a partial dephasing CPTP channel  ${\cal D}_n$  which is more destructive than  $\mathcal{A}_{\infty,n}$ as it only preserves coherence into the subspace formed by the first $n-1$ Fock states while inducing dephasing on the remaining ones (its 
Kraus operators being hence  $\hat{P}_{< n}$ plus the collection of  one-dimensional projectors on each of the higher-photon-number states). 
In Fig.~{\ref{kenCompare} we have tested the performance of this new detection scheme for the case $n=2$ observing only  a tiny decrement (of order $O(10^{-4})$)  of the associated
success probability  with respect to the one obtained in  the nh-P-Amp case, proving hence that, rather counterintuitively, the dephasing transformation  ${\cal D}_n$ is a useful resource for the discrimination problem we are considering here. 
An implementation of ${\cal D}_2$ can be obtained by 
sending the incoming coherent signal $|\alpha\rangle$ inside an optical cavity, coupled to a two-level atom initialized into its ground state $\ket{G}$, via the Jaynes-Cummings hamiltonian \cite{NChuang}
\begin{eqnarray}
\hat{H}_{JC}=\omega \hat{N}+\gamma(\hat{a}^{\dag}\hat{\sigma}_{-}+\hat{a}\hat{\sigma}_{+})\label{ham},
\end{eqnarray} 
where $\hat{N}=\hat{a}^{\dag}\hat{a}+\hat{Z}/2$ is a first integral of motion, obtained by combining the cavity bosonic creation and annihiliation operators $\hat{a}^{\dag}$, $\hat{a}$ and the atomic energy operator $\hat{Z}=\ketbra{E}-\ketbra{G}$, with $\ket{E}$  the excited atomic state; $\omega$ is the frequency of both cavity and atom at resonance. Finally the second term entering $H_{JC}$ represents the cavity-atom coupling of strength $\gamma$, where $\hat{\sigma}_{\pm}$ are the atomic operators describing excitation and decay of its quantum state. 
To induce the transformation ${\cal D}_2$,  we first let the coherent signal and the atom interact for a time $\tau$ chosen in such a way to induce a perfect Rabi oscillation. This guarantees that the joint cavity-atom state $\ket{1,G}$ of the input superposition $\ket{\alpha,G}$ is transferred to $\ket{0,E}$, thus encoding the zero and one-photon-number cavity subspace in the atomic levels. This happens for the first time at $\tau=\pi/(2\gamma)$, leaving the system in the joint state
\begin{eqnarray}
|\psi_{RABI}\rangle = e^{-\frac{\norm{\alpha}^{2}}{2}}( \ket{0,G}-i\alpha e^{-i\omega \tau}\ket{0,E} + |\Delta \rangle),   \label{firstRabi} 
\end{eqnarray}
with $|\Delta \rangle$ being a  combination of terms which, on the optical part, posses at least one photon excitation. 
 Next we abruptly decouple the two systems (say detuning the atom energy gap  with respect to the cavity frequency) while inducing a random perturbation on the cavity wavelength. Alternatively,
 we may assume  the optical signal to emerge from the cavity and to be  fed into a long waveguide that dephases the various Fock components of the propagated signals. 
 In both cases the net effect on  $|\psi_{RABI}\rangle$ can be described as an application of the operator  $\exp(-i\theta\hat{a}^{\dag}\hat{a})$ with $\theta$ being a random parameter we have to average over, 
 while  no phase is added to the first two components of the global state \eqref{firstRabi}, containing the superposition we want to preserve.
After this stage  we apply a second Rabi oscillation in order to bring back the preserved atomic superposition onto the cavity states (e.g. by abruptly restoring the atom-field resonance condition or by
feeding the traveling signal back to the cavity). We describe this process with the same Jaynes-Cummings hamiltonian and interaction time as before. The output field of the cavity, obtained by tracing out the atomic state and averaging over the random phase, can be written as
\begin{eqnarray} 
\rho =e^{-\norm{\alpha}^{2}}\Big[\ketbra{\alpha_{T}}+\frac{\norm{\alpha}^{4}}{2}\left(D\ketbra{1}+\alpha E\dketbra{2}{1}+\text{h.c.}\right) \nonumber \\ \nonumber 
+\sum_{k=2}^{\infty}\frac{\norm{\alpha}^{2k}}{k!}\left(D_{k}(\alpha)\ketbra{k}+\alpha E_{k}(\alpha)\dketbra{k+1}{k}+\text{h.c.}\right)\Big],
\end{eqnarray} 
where $\ket{\alpha_{T}}=\ket{0}-\alpha \exp(-2i\omega\tau)\ket{1}$ is the superposition we aimed at preserving, apart from a phase which can be dropped out either by fine tuning the working frequency $\omega$ or by earlier compensation of the input coherent states; the various $D$, $E$ coefficients assume non-trivial values, having fixed $\tau$ in order to favor the desired Rabi transitions. The main imperfection of this implementation with respect to the desired partial dephaser ${\cal D}_2$ 
is that  the final state $\rho$  actually preserves some extra coherence between adjacent photon-number states. Nevertheless, if we employ this device instead of the ideal dephaser in our receiver setup, the top increase of the success probability previously obtained with the nh-P-Amp at $\alpha\simeq0.32$ is reduced to $\sim1.58\%$, i.e. we have a performance loss of only up to $0.27\%$ in the low-intensity region $\norm{\alpha}^{2}\lesssim 0.1$, see Fig. \ref{kenCompare}. Such loss increases considerably at higher intensities, where the nh-P-Amp itself provides a smaller advantage over optimized Kennedy detection, but this features can be easily superseded by performing a few multiplexing steps in a Dolinar-like fashion (compare Fig. \ref{kenCompare} and \ref{dol2Step}).
 \begin{figure}[h!]
\includegraphics[scale=.265]{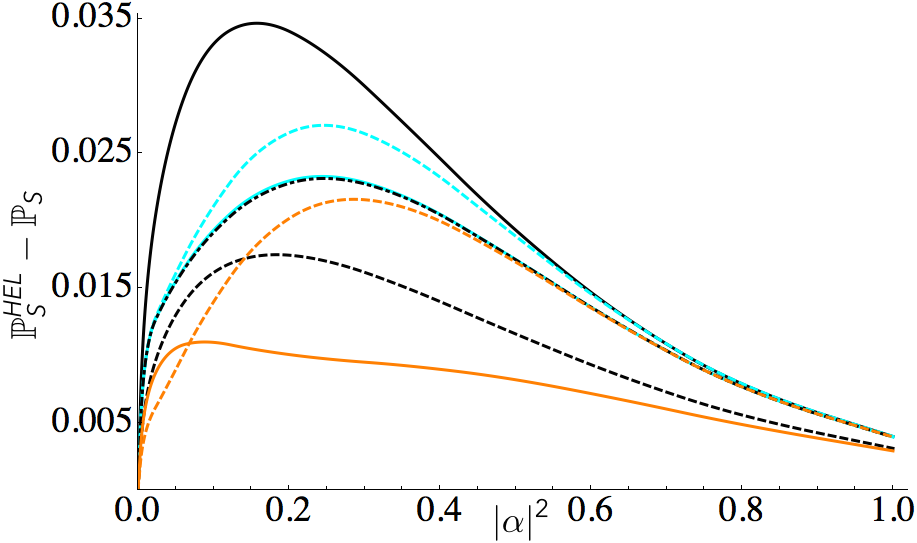}
\caption{(Colour online) Plot of the difference between the success probability of  Helstrom and the one of several Kennedy-like receivers, as a function of the input states' average photon number~$\norm{\alpha}^{2}$: optimized Kennedy scheme (black solid line), ${\cal A}_{\infty,2}$ scheme (black dot-dashed line), ${\cal D}_{2}$ scheme (cyan/light gray solid line), cavity implementation  (cyan/light gray dashed line), Takeoka-Sasaki (TS) scheme~\cite{OpGauss} with squeezing and displacement (black dashed line), ${\cal A}_{\infty,2}$ plus TS scheme (orange/dark gray dashed line), ${\cal A}_{\infty,3}$ plus TS scheme (orange/dark gray solid line). 
 }\label{kenCompare}
\end{figure}

\section{Combination with active gaussian unitary} \label{sec4}
A possible extension of our scheme
can be obtained along the line proposed in Ref.~\cite{OpGauss} where a standard Kennedy scheme has been improved by adding  a squeezing operation $\hat{S}_{r}=\exp((\hat{a}^{2}-\hat{a}^{\dag 2})r/2)$ of parameter $r\in\mathbb{R}$  before the last  displacement transformation, 
resulting in an overall active (non phase-insensitive) gaussian unitary before photon detection. 
When we applied the same method to our scheme the   probabilities~(\ref{OVERLAP1}) and (\ref{OVERLAP}) get replaced by
\begin{eqnarray}
P(0|\alpha_{0})&=&\norm{\dbraket{\beta,-r}{0}}^{2}\;, \\
P(0|\alpha_{1})&=&\langle \beta, -r | \mathcal{A}_{g,n}(\ketbra{2\alpha})|\beta,-r\rangle\;,
\end{eqnarray} 
where  $\ket{\beta,-r}=\hat{S}_{-r}\hat{D}_{\beta}\ket{0}$ is a displaced-squeezed state. 
We can optimize the associated success probability \eqref{succprobAmp} with respect to both $\beta$ and $r$. At variance with the passive-gaussian scheme, setting $n=2$ here turns out to be optimal only in the low-intensity region $\norm{\alpha}^{2}\lesssim 0.1$. For higher intensity values, an amplifier cutoff $n=3$ attains instead the most satisfying results, clearly surpassing all other receivers in performance (Fig. \ref{kenCompare}); this is probably due to the fact that the squeezing operation requires additional coherence terms between the zero and one-photon subspace and the two-photons one in order to be properly optimized. In particular the optimal squeezing is found at negative values of the parameter $r<0$ implying 
hence a squeezing of the $\hat{p}$ quadrature.

\section{Comparison with the Dolinar receiver} \label{sec5}
Since the proposed detection schemes only require the insertion of an additional operation in a Kennedy-like receiver, it seems reasonable to study their extension to a Dolinar-like one, in the multiplexed version proposed in \cite{ImplProj,Multicopy,DolMultiplexed}. Accordingly we  now preliminary map
the input coherent states $\ket{\pm\alpha}$  into $N$ low-intensity copies 
$\ket{\pm\alpha/\sqrt{N}}^{\otimes N}$ which we probe  in sequence exploiting the information acquired at each stage to optimize the 
 parameters (e.g. displacement, amplification, cutoff) of the detection that follows. 
 In Fig.~\ref{dol2Step} we show the success probability of Dolinar-like detection schemes, for the simplest case $N=2$, taking the Helstrom bound as a reference. As it may be expected, since the proposed schemes outperform the Kennedy receiver, they also outperform the Dolinar one. The inset shows the same quantity as a function of the number of steps $N$ and fixed intensity. 
 \begin{figure}[t]
\includegraphics[scale=.46]{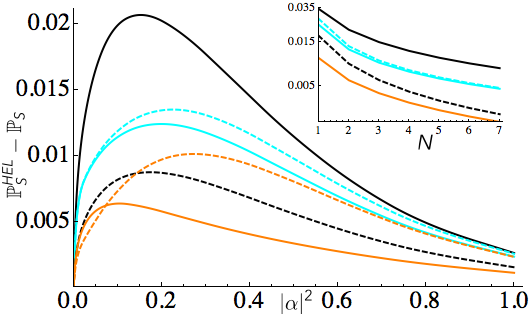}
\caption{(Colour online) Plot of the difference between the success probability of Helstrom and that of  several Dolinar-like protocols, as a function of the input states average photon number~$\norm{\alpha}^{2}$, for two multiplexing steps: simple Dolinar scheme with gaussian optimized displacement (black solid line), ${\cal D}_2$ scheme (cyan/light gray solid line), cavity implementation  (cyan/light gray dashed line), TS scheme, with gaussian squeezing and displacement (black dashed line), ${\cal A}_{\infty,2}$  plus TS scheme (orange/dark gray dashed line),  ${\cal A}_{\infty,3}$ plus TS scheme (orange/dark gray solid line). 
The inset shows the same quantity (log-scale) as a function of the number of steps $N$ at fixed $\norm{\alpha}^{2}=0.2$, for the same protocols as in the main picture (only 
the case ${\cal A}_{\infty,3}$ is plotted for the dephaser plus TS scheme).}\label{dol2Step}
\end{figure}
 \section{Conclusions} \label{sec6}
  We have analyzed the performance of the non-linear nh-P-Amp in a binary coherent-state discrimination task. Our contribution is twofold. 
  On one hand it is the first time, to our knowledge, that the P-Amp is being employed regardless of its probabilistic nature, which so far constituted a major drawback to its use as an amplifier. In particular the non-heralding procedure allows us to recycle the highly-probable non-amplified states and perform a conclusive discrimination in any case. While the resulting improvements are not particularly high, 
the proposed scheme
   appears to be beneficial  
 also  when properly coupled to any of the currently used instruments: gaussian transformations, single-photon detectors, adaptive operations. In particular, it seems to perform well in further refining detection protocols based on signal nulling. 
On the other hand, we point out that performances analogous to those observed for the nh-P-Amp scheme can also be obtained
when replacing the latter with a partial dephasing channel which preserves  coherence among  low energy vectors.  We find this last observation conceptually intriguing as dephasing transformations are noisy transformations and hence typically associated with detrimental, not useful, effects.

 \appendix
\section{The  necessity of a second displacement} \label{APPA}
As discussed in the main text, our decoding scheme accounts for three steps to be performed prior of the photo-detection. Specifically: 
{\it i)}  an initial nullifying displacement which brings $\alpha_0$ into the vacuum and $\alpha_1$ into $2\alpha$; {\it ii)}  the application of ${\cal A}_{g,n}$;
{\it iii)} a final displacement of~$-\beta$. 
Here we show that the latter step is fundamental to get an improvement with respect to the standard
Kennedy scheme (the latter being described by dropping also the amplification from the list). 
To see this we observe that the probabilities entering~(1) can be expressed as the following Uhlmann 
 fidelities~\cite{NChuang}   
$P(0|\alpha_{0})=F(\ketbra{\beta},\ketbra{0})$,  
$P(0|\alpha_{1})=F(\ketbra{\beta},\mathcal{A}_{g,n}(\ketbra{2\alpha}))$.
Similarly for the Kennedy scheme we have 
$P_{KEN}(0|\alpha_{0}) = F(\ketbra{\beta},\ketbra{0})$, 
$P_{KEN}(0|\alpha_{1})=  F(\ketbra{\beta},\ketbra{2\alpha})$.
Accordingly, while the vacuum-overlap of the amplified favored state is always trivially equal to its optimized Kennedy counterpart, no general ordering 
can be found for the other state, unless the final displacement is set to zero; indeed for $\beta=0$ we can use the fact that ${\cal A}_{g,n}$ leaves the vacuum invariant and  the non-decreasing property of fidelity under CPTP evolution of both its arguments to show that 
\begin{eqnarray}
P(0|\alpha_{1})&=&F(\ketbra{0},\mathcal{A}_{g,n}(\ketbra{2\alpha})) \nonumber \\
&=& F(\mathcal{A}_{g,n}(\ketbra{0}),\mathcal{A}_{g,n}(\ketbra{2\alpha})) \nonumber \\ &\geq& 
 F(\ketbra{0},\ketbra{2\alpha}) = P_{KEN}(0|\alpha_{1}),
\end{eqnarray}
and hence 
${\mathbb P}_{{\bf S}}(\alpha,g,n,0)\leq  {\mathbb P}^{KEN}_{{\bf S}}(\alpha,0) \leq {\mathbb P}_{{\bf S},opt}^{KEN}(\alpha)$.
\section{Ordinary parametric amplification does not help} \label{APPB}
Here we show that by replacing ${\cal A}_{g,n}$ with an ordinary parametric amplifier or, more generally, a phase-insensitive gaussian channel $\Phi$ does not improve detection. Indeed such a channel can be always decomposed \cite{PhInsGauss,PhInsGauss2} as $\Phi=\mathcal{A}_{k}\circ\mathcal{E}_{\eta}$, i.e. the concatenation of a quantum-limited attenuator of parameter $\eta\leq 1$ and amplifier of parameter $k\geq 1$. By noticing that the latter has a dual channel $\mathcal{A}_{k}^{*}=k^{-1}\mathcal{E}_{k^{-1}}$, we can write the overlap between coherent states $\ket{\alpha}$ and $\ket{\beta}$ after transformation of the former under $\Phi$ as 
\begin{eqnarray} 
&&F\left(\ketbra{\beta}),\Phi(\ketbra{\alpha})\right)=k^{-1}F\left(\mathcal{E}_{k^{-1}}^{*}(\ketbra{\beta}),\mathcal{E}_{\eta}(\ketbra{\alpha})\right) \nonumber \\
&&\qquad \qquad \qquad  \qquad \qquad =k^{-1}F\left(\ketbra{\beta'},\ketbra{\sqrt{\eta}\alpha}\right), \nonumber 
\end{eqnarray} 
$\ket{\beta'}$ being an attenuated version  of $\ket{\beta}$ whose explicit value is irrelevant since it will be optimized. Thus, calling $\Delta_{\Phi,\alpha}$, $\Delta_{\alpha}$ the difference between vacuum-overlaps of the two input states respectively with and without application of the channel $\Phi$, we have $\Delta_{\Phi,\alpha}=k^{-1}\Delta_{\sqrt{\eta}\alpha}\leq\Delta_{\alpha}$, i.e. no improvement in the success probability can be obtained by applying a phase-insensitive gaussian channel before detection.

\end{document}